\documentclass[aps,twocolumn,amsmath,amssymb]{revtex4-2}
\usepackage[dvipdfmx]{graphicx} 
\usepackage{bm}
\usepackage{braket}
\usepackage{hyperref}
\hypersetup{colorlinks=true, urlcolor=blue, linkcolor=blue, citecolor=blue}

\begin{document}

\title{Correlation-driven branch in doped excitonic insulators}

\author{Tatsuya Kaneko,$^1$ Ryota Ueda,$^1$ and Satoshi Ejima$^2$}

\affiliation{$^1$Department of Physics, The University of Osaka, Toyonaka, Osaka 560-0043, Japan \\
$^2$Institut f\"ur Softwaretechnologie, Abteilung High-Performance Computing, Deutsches Zentrum f\"ur Luft- und Raumfahrt (DLR), 22529 Hamburg, Germany} 

\date{\today}

\begin{abstract}
We investigate the spectral properties of doped one-dimensional excitonic insulators. 
Employing matrix-product-state-based methods, we compute the single-particle spectrum and optical conductivity in a correlated two-band model. 
Our numerical calculation reveals the emergence of a correlation-driven in-gap branch in the doped state. 
The origin of the in-gap branch is examined by decomposing the propagation dynamics of a single particle, elucidating that the doping-induced branch is associated with excitonic correlations. 
Our demonstrations suggest that the doping-induced branch can serve as an indicator of electron-hole correlations. 
\end{abstract}

\maketitle


\section{Introduction}

Understanding many-body properties in correlated systems is a central issue in condensed matter physics~\cite{MImada1998,EDagotto2005}. 
Excitonic insulators (EIs) are correlation-driven insulators in multiband systems~\cite{LKeldysh1965,JDescloizeaux1965,DJerome1967,BHalperin1968,TKaneko2025}. 
In a semiconductor with a narrow gap or a semimetal with a small band overlap, interband Coulomb interactions modify its electronic structure, and the resulting gap opening leads to an EI state. 
To date, various candidate materials for EIs have been proposed by both theory and experiment~\cite{MTraum1978,HCercellier2007,AKogar2017,QGao2024,YWakisaka2009,KSeki2014,TYamada2016,YLu2017,KKim2021,KFukutani2021,JHuang2024,PZhang2024,MHossain2025,JKunes2014_2,AIkeda2023,DVarsano2020,YJia2022,BSun2022,BBucher1991,DMazzone2022,HQu2025,ROkuma_arXiv}. 
In several candidate materials that undergo structural transitions, such as TiSe$_2$ and Ta$_2$NiSe$_5$~\cite{FDiSalvo1976,FDiSalvo1986}, electron-lattice coupling can contribute to the deformations of their electronic structures~\cite{JvanWezel2010,CMonney2012,TKaneko2013,BZenker2014,YMurakami2020}, giving rise to controversy over the origin of their phase transitions~\cite{ZLin2022,EBaldini2023,CChen2023}. 
Given this background, detecting the excitonic contribution to the ground-state configuration is an important challenge in the study of EIs and their candidate materials. 

Carrier doping can be used to assess the contribution of correlation effects to the ground-state configuration because band insulators and correlation-driven insulators exhibit distinct behaviors with respect to doping. 
In noninteracting band insulators where the independent-particle picture is valid, carrier doping can be achieved by shifting the Fermi level without deforming the band structure. 
However, this rigid-band picture is often inapplicable to strongly correlated systems. 
For example, hole doping into Mott insulators induces changes in the structure of the single-particle spectrum, where the spectral weight transfer from the upper Hubbard band to near the Fermi level and the emergence of the doping-induced in-gap branch, reflecting a spin excitation, have been revealed by numerical calculations~\cite{HEskes1991,EDagotto1992,MMeinders1993,NBulut1994,RPreuss1995,DSenechal2000,BKyung2006,MKohno2010,MKohno2012}. 
Since the EI state is also a correlation-driven insulating state, doping-induced band deformation, which is unexpected in conventional band insulators, may occur in EIs. 

In this paper, we investigate the effects of carrier doping on a one-dimensional (1D) EI state described in a correlated two-band model. 
We compute the single-particle spectrum and optical conductivity employing matrix-product-state-based methods. 
We find the emergence of a doping-induced in-gap branch in the single-particle spectrum. 
In the optical conductivity, the original peak structure based on interband transitions is maintained, while the Drude weight grows with doping, reflecting the structure of the single-particle spectrum. 
The decomposition of the single-particle dynamics reveals that the doping-induced branch is associated with excitonic correlations. 

The rest of this paper is organized as follows. 
In Sec.~\ref{sec:model&method}, we introduce our target system, the spectral functions, and the numerical techniques employed in this study. 
In Sec.~\ref{sec:results}, we present the calculated single-particle spectrum, optical conductivity, and excitonic correlation function. 
A summary of our study is given in Sec.~\ref{sec:summary}.


\section{Model and Method} \label{sec:model&method}

\subsection{Model}

To describe an EI state, we employ the simplest correlated two-band model~\cite{CBatista2002,DIhle2008,KSeki2011,SMor2017,KSugimoto2018,TTanabe2021,AOsterkorn2025} whose 1D Hamiltonian is given by 
\begin{align}
\hat{H} = 
&-\sum_{j} \sum_{\alpha} t_{\alpha} \left( \hat{c}^{\dag}_{j,\alpha} \hat{c}_{j+1,\alpha} + {\rm H.c.} \right)
\notag \\
&+\frac{D}{2} \sum_{j} \left( \hat{n}_{j,a} - \hat{n}_{j,b} \right) 
+U \sum_{j} \hat{n}_{j,a} \hat{n}_{j,b} .
\label{eq:Hamiltonian}
\end{align}
$\hat{c}^{\dag}_{j,\alpha}$ ($\hat{c}_{j,\alpha}$) is the creation (annihilation) operator of a spinless fermion in orbital $\alpha$~$(=a,b)$ at site $j$. 
$t_{\alpha}$ is the hopping parameter between nearest-neighboring sites on orbital $\alpha$. 
$D$ is the energy level difference between two orbitals. 
$U$ is the interorbital Coulomb interaction. 
We consider a direct-gap type band structure with $t_a t_b < 0$. 
In this paper, $t_a=t_{\rm h}$ ($>0$) is the unit of energy. 
We set $t_b=-t_a$ unless otherwise specified. 
Throughout the paper, the Planck constant ($\hbar$), the charge of a particle ($e$), and the lattice constant are set to 1. 
The number of lattice sites and the number of fermions are denoted by $L$ and $N$, respectively. 
$N = N_a + N_b$, where $N_{\alpha}$ is the number of fermions in orbital $\alpha$. 
We address the case where $D>0$ and $N_b > N_a$. 
When $U=0$ without doping, the $a$-orbital chain forms the conduction band (CB), and the $b$-orbital chain forms the valence band (VB). 
We consider hole doping away from half filling ($N=L$) and define the doping concentration as $\delta = 1 - N/L$.

\subsection{Spectral functions}

We compute the single-particle spectrum to visualize the band structure in the correlated model. 
The single-particle spectrum $A(k,\omega)$ is obtained by the Fourier transformation of the retarded Green's function $G^{\rm R}(i-j,t-t') = -i \theta(t-t') \sum_{\alpha} \braket{\psi_0 | \{ \hat{c}_{i,\alpha}(t), \hat{c}^{\dag}_{j,\alpha}(t') \} |\psi_0}$, where $\ket{\psi_0}$ is the ground state, $\theta(t)$ is the step function, and $\hat{\mathcal{O}}(t) = e^{i\hat{H}t} \hat{\mathcal{O}} e^{-i\hat{H}t}$ for an operator $\hat{\mathcal{O}}$. 
In our numerical calculations using finite-size systems, the following two functions are computed: 
\begin{align}
&G^{+}_{\alpha}(x,t) = - i \Braket{\psi_0 | \hat{c}_{j_0+x,\alpha}(t) \hat{c}^{\dag}_{j_0,\alpha}(0) |\psi_0}, 
\\
&G^{-}_{\alpha}(x,t) = - i \Braket{\psi_0 | \hat{c}^{\dag}_{j_0,\alpha}(0) \hat{c}_{j_0+x,\alpha}(t) |\psi_0}, 
\end{align}
where $t>0$ and $j_0=L/2$. 
$G^{+}_{\alpha}$ ($G^{-}_{\alpha}$) describes the dynamical properties of a fermion above (below) the Fermi level $E_{\rm F}$. 
The Fourier transformation of $G_{\alpha}(x,t) = G^{+}_{\alpha}(x,t) + G^{-}_{\alpha}(x,t)$ is conducted by 
\begin{align}
G_{\alpha}(k,\omega) = \sum_x e^{-ikx} \int^{\infty}_0 dt \, e^{i(\omega + i\eta)t} G_\alpha(x,t). 
\end{align}
$\eta$ is introduced to perform numerical integrations using data from finite-time simulations up to $t_{\rm max}$ instead of $t \rightarrow \infty$. 
The single-particle spectrum of component $\alpha$ is given by 
\begin{align}
A_{\alpha} (k,\omega) = -\frac{1}{\pi} {\rm Im} G_{\alpha}(k,\omega) . 
\end{align}
The total single-particle spectrum is given by $A(k,\omega)= \sum_{\alpha}A_{\alpha} (k,\omega)$. 

We also consider the optical conductivity $\sigma(\omega)$. 
In this paper, the optical conductivity is computed by 
\begin{align}
\sigma (\omega) = \frac{\kappa_{T} + \chi_{JJ}(\omega)}{i(\omega+i\eta)}. 
\end{align}
$\kappa_{T} = \braket{\psi_0 | \hat{T} | \psi_0}/L$ with $\hat{T} = -\sum_{j,\alpha} t_{\alpha} ( \hat{c}^{\dag}_{j,\alpha} \hat{c}_{j+1,\alpha} + {\rm H.c.} )$ corresponds to the kinetic energy. 
$\chi_{JJ}(\omega)$ is the dynamical current-current correlation function 
\begin{align}
\chi_{JJ} (\omega) = 
\frac{i}{L} \int^{\infty}_0 dt \, e^{i(\omega + i\eta)t} \Braket{\psi_0 | \left[ \hat{J}(t), \hat{J}(0) \right] | \psi_0},
\end{align}
where $\hat{J} = - i \sum_{j,\alpha} t_{\alpha} ( \hat{c}^{\dag}_{j,\alpha} \hat{c}_{j+1,\alpha} - {\rm H.c.} )$ is the current operator.

\subsection{Method}

We employ the density-matrix renormalization group (DMRG) method~\cite{SWhite1992,SWhite1993,USchollwock2011} to obtain the ground state $\ket{\psi_0}$. 
The ground-state properties of the nondoped insulating state have been investigated in a previous DMRG calculation in Ref.~\cite{SEjima2014}. 
We use the 1D chain of $L=160$ sites with open boundary conditions. 
In the present model, $M = N_b - N_a$ can be a quantum number. 
We evaluate the energy as a function of $M$ and employ the state with the lowest energy as the ground state. 
The bond dimension for the ground-state calculation is up to $m=2000$, where the largest truncation error is less than $10^{-10}$. 
We evolve the system in real time using the time-evolving block decimation (TEBD) algorithm~\cite{GVidal2003} to obtain the dynamical correlation functions. 
The time evolution is carried out using the second-order Suzuki-Trotter decomposition~\cite{SPaeckel2019}. 
The time step $\delta t = 0.01/t_{\rm h}$ and the bond dimension $m=1000$ are used for the time evolution up to $t_{\rm max} = 40/t_{\rm h}$. 
The damping factor, which gives Lorentzian broadening in the spectral functions, is $\eta=0.125t_{\rm h}$. 
We employ the window function $w(t) = [ 1 + \cos(\pi t/t_{\rm max}) ]/2$ in the Fourier transformation to reduce artifacts from the finite-time cutoff at $t_{\rm max}$~\cite{SMilner2026}. 
Unless otherwise specified, $A_{\alpha}(k,\omega)$ is plotted in units of $t_{\rm h}^{-1}$. 
The upper~($+$) and lower~($-$) edges of the band gap of the insulating state are given by $\pm [ E_{\rm min}(N \pm 1) - E_0(N) ]$, where $E_0(N)$ is the ground-state energy of $N$ fermions and $E_{\rm min}(N\pm 1)$ is the minimum energy when one fermion is added ($+1$) or removed ($-1$) from the ground-state configuration~\cite{Emin}. 
The Fermi level is defined as $E_{\rm F} = [ E_{\rm min}(N+1) - E_{\rm min}(N-1)]/2$.


\section{Results} \label{sec:results}

\subsection{Single-particle spectrum}

\begin{figure}[t]
\begin{center}
\includegraphics[width=\columnwidth]{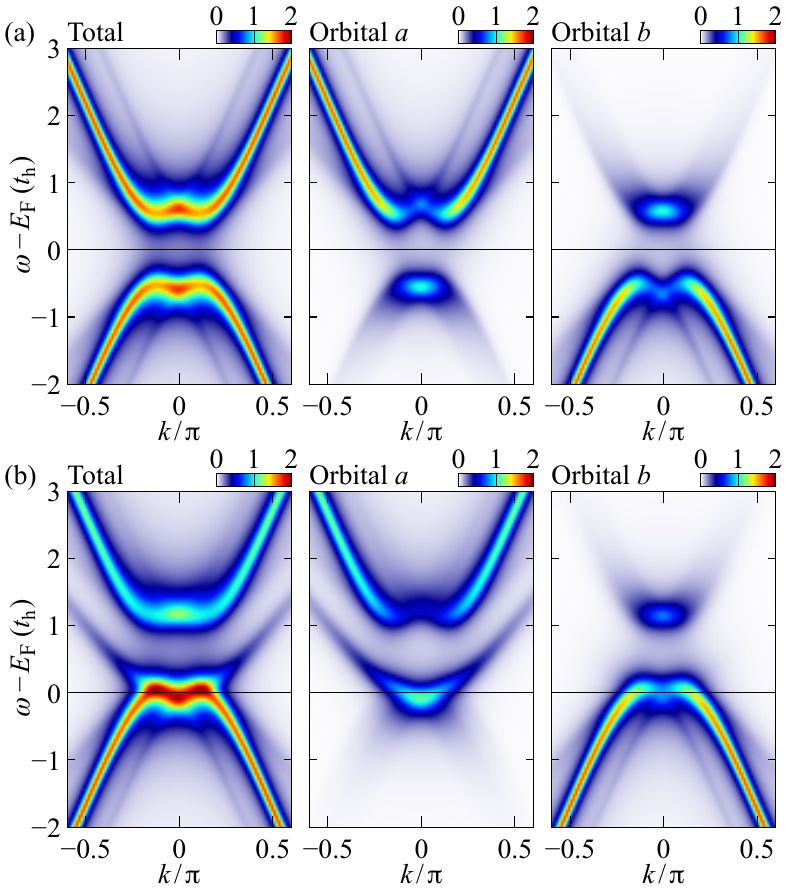} 
\caption{Single-particle spectra for (a) $\delta=0$ (half-filling) and (b) $\delta=0.05$ (hole doping), where $U=3t_{\rm h}$ and $D=1.95t_{\rm h}$. 
The left, middle, and right panels display $A(k,\omega)$ (total), $A_a(k,\omega)$ (orbital $a$), and $A_b(k,\omega)$ (orbital $b$), respectively. 
The horizontal lines represent the Fermi level $E_{\rm F}$.} 
\label{fig1}
\end{center}
\end{figure}

\begin{figure}[t]
\begin{center}
\includegraphics[width=\columnwidth]{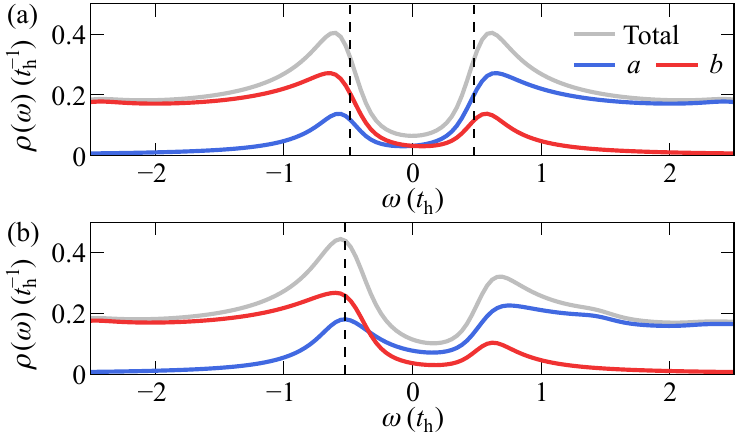} 
\caption{Densities of states for (a) $\delta=0$ (half filling) and (b) $\delta=0.05$ (hole doping), where $U=3t_{\rm h}$ and $D=1.95t_{\rm h}$. 
The energy of the horizontal axis is shifted by $U/2$, where the center of the gap is at zero in (a). 
The vertical dashed lines in (a) indicate the upper and lower edges of the band gap. 
The vertical dashed line in (b) represents the Fermi level $E_{\rm F}$.} 
\label{fig2}
\end{center}
\end{figure}

Figure~\ref{fig1}(a) presents the single-particle spectra of the nondoped state at $U=3t_{\rm h}$. 
In the noninteracting case ($U=0$), the CB and VB, which are formed by the $a$ and $b$ orbitals, respectively, overlap around the $\Gamma$ ($k=0$) point. 
However, as shown in Fig.~\ref{fig1}(a), the band gap opens, and the band edges become flat near the $\Gamma$ point due to the Coulomb repulsion $U$. 
On the flat structure near the $\Gamma$ point, the spectral weight of the $a$ ($b$) orbital remains below (above) $E_{\rm F}$. 
This insulating state is purely due to the interband (i.e., excitonic) interaction $U$, and we regard this state as an EI state in a 1D system. 
The spectral properties of the nondoped state are consistent with the previous studies~\cite{RFujiuchi2019,SEjima2021,SEjima2022}. 

The single particle spectra of the doped state are shown in Fig.~\ref{fig1}(b). 
In addition to the downward shift of $E_{\rm F}$, the spectral structure itself is deformed from the nondoped state, indicating that the rigid-band picture is not valid. 
While the original gap structure due to $U$ seen in the nondoped state remains, the in-gap branch emerges from the lower band, and the doped state becomes metallic. 
As shown in $A_a(k,\omega)$ [middle panel of Fig.~\ref{fig1}(b)], the doping-induced branch stems from the spectral weight of the $a$-orbital component and develops above $E_{\rm F}$. 

As a result of doping, the spectral weight in the upper band decreases and is redistributed to the region near $E_{\rm F}$. 
To clarify the doping-induced spectral-weight transfer, we plot the density of states (DOS) $\rho_{\alpha}(\omega) = (1/L) \sum_{k} A_{\alpha}(k,\omega)$ in Fig.~\ref{fig2}, where $\omega=0$ is set at the center of the gap in the nondoped state ($\delta=0$). 
Note that the nonzero weight in the gap of the insulating state at $\delta=0$ [Fig.~\ref{fig2}(a)] is due to the broadening factor $\eta$ introduced in the numerical calculation. 
Compared with the nondoped insulating state [Fig.~\ref{fig2}(a)], the total DOS in the upper band is suppressed [Fig.~\ref{fig2}(b)], while the spectral weights near $E_{\rm F}$ increase in the doped state. 
This spectral property is similar to that of Mott insulators~\cite{HEskes1991,MMeinders1993}. 
Hence, carrier doping in the 1D EI modifies the structure of the single-particle spectrum, distinguishing it from carrier doping in band insulators described in the rigid-band picture.

\subsection{Optical conductivity}

\begin{figure}[t]
\begin{center}
\includegraphics[width=\columnwidth]{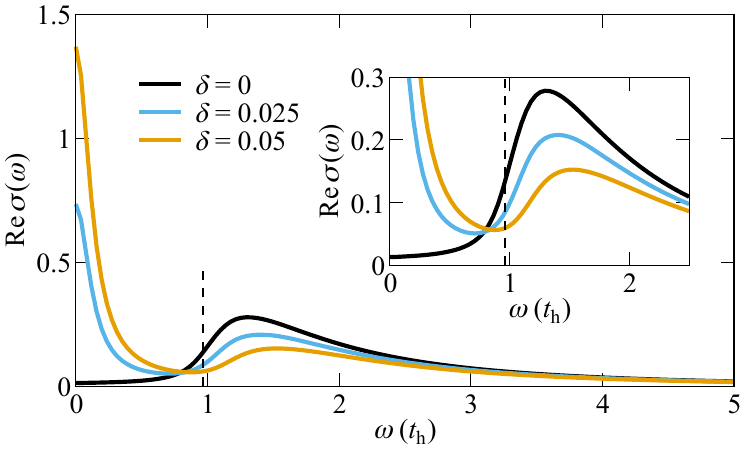} 
\caption{Real part of the optical conductivity $\sigma(\omega)$ for various $\delta$, where $U=3t_{\rm h}$ and $D=1.95t_{\rm h}$. 
The vertical dashed line represents the single-particle excitation gap of the nondoped insulator ($\delta=0$). 
Inset: Enlarged view of the spectra. 
Note that when $\delta=0$, the nonzero optical conductivity for $\omega$ below the gap is due to the broadening factor introduced in the numerical calculation.} 
\label{fig3}
\end{center}
\end{figure}

Figure~\ref{fig3} presents the real part of the optical conductivity $\sigma(\omega)$ at $U=3t_{\rm h}$. 
The spectral weight appears above the single-particle gap when $\delta=0$ (dashed line in Fig.~\ref{fig3})~\cite{DE}, whereas doping changes the structure of the optical conductivity. 
First, the peak at $\omega=0$, corresponding to the Drude weight, grows as the doping $\delta$ increases, reflecting the metallic nature of the doped state, which is consistent with the single-particle spectrum shown in Fig.~\ref{fig1}(b). 
Second, the peak structure above the gap when $\delta=0$ (dashed line in Fig.~\ref{fig3}) remains even with hole doping, while its spectral weight decreases. 
The remnants of the gap structure are consistent with the single-particle spectral features of the doped state shown in Fig.~\ref{fig1}(b), indicating that the optical conductivity $\sigma(\omega)$ captures both the metallic nature and the band structure of the doped EI.

\subsection{Origin of the doping-induced branch}

\begin{figure}[t]
\begin{center}
\includegraphics[width=\columnwidth]{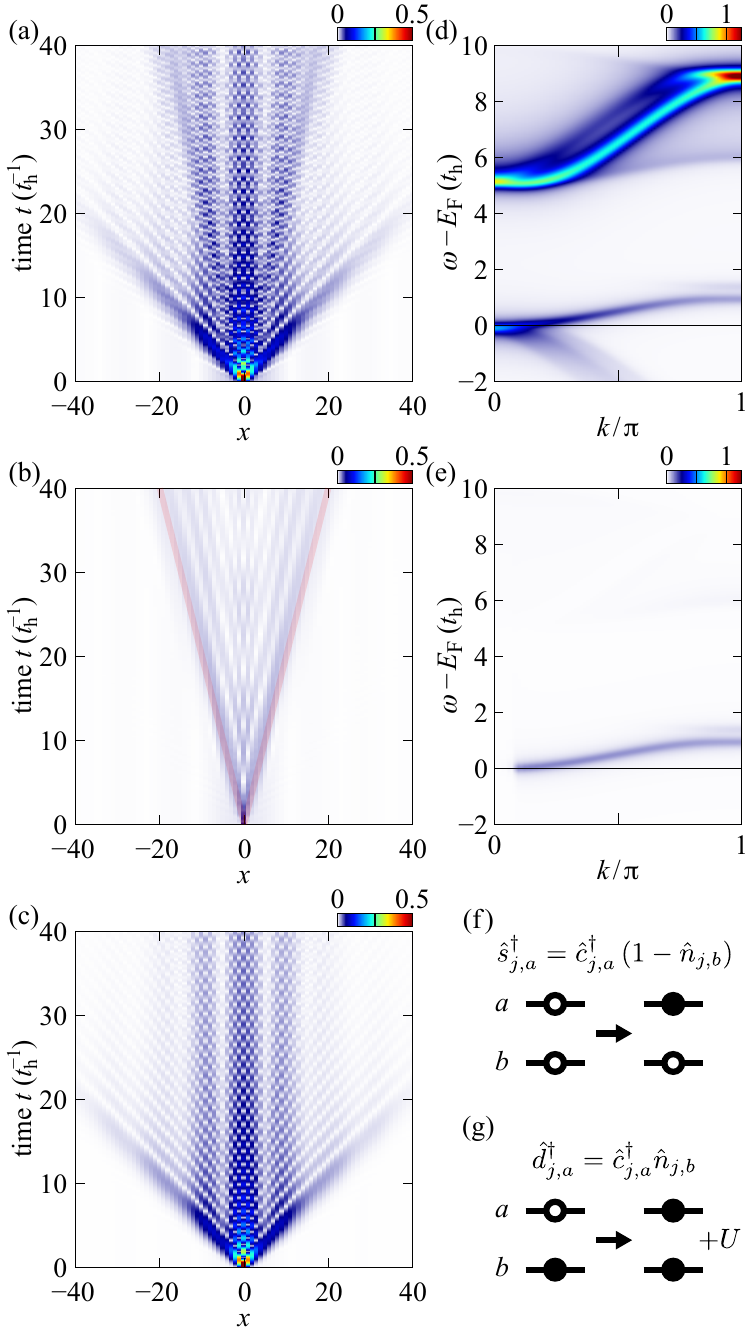} 
\caption{Absolute values of the space-time correlation functions $\mathcal{G}_a(x,t) = -i \braket{\psi_0 | \hat{\mathcal{O}}_{j+x,a}(t) \hat{\mathcal{O}}^{\dag}_{j,a}(0) | \psi_0}$ for (a) $\hat{\mathcal{O}}^{\dag}_{j,a} = \hat{c}^{\dag}_{j,a}$, (b) $\hat{\mathcal{O}}^{\dag}_{j,a} = \hat{s}^{\dag}_{j,a}=\hat{c}^{\dag}_{j,a} (1-\hat{n}_{j,b})$, and (c) $\hat{\mathcal{O}}^{\dag}_{j,a} = \hat{d}^{\dag}_{j,a}=\hat{c}^{\dag}_{j,a} \hat{n}_{j,b}$, where $U=8t_{\rm h}$, $D=0.92t_{\rm h}$, and $\delta=0.05$. 
The translucent red lines in (b) represent $x=\pm J_{\rm ex}t$, where $J_{\rm ex}=4t_at_b/U$. 
(d) Single-particle spectrum $A_{a}(k,\omega)$ and (e) spectrum obtained by the Fourier transformation of $\mathcal{G}_a(x,t)$ for $\hat{s}^{\dag}_{j,a}$ in (b). 
Schematic figures of the operators (f) $\hat{s}^{\dag}_{j,a} = \hat{c}^{\dag}_{j,a} (1-\hat{n}_{j,b})$ and (g) $\hat{d}^{\dag}_{j,a} = \hat{c}^{\dag}_{j,a} \hat{n}_{j,b}$.} 
\label{fig4}
\end{center}
\end{figure}

Let us now discuss the origin of the doping-induced branch seen in Fig.~\ref{fig1}(b). 
Since the branch grows above $E_{\rm F}$ in the $a$-orbital component [middle panel of Fig.~\ref{fig1}(b)], we consider $G^{+}_{a} (x,t)= - i \braket{\psi_0 | \hat{c}_{j+x,a}(t) \hat{c}^{\dag}_{j,a}(0) | \psi_0}$, which describes the propagation of a created fermion along the $a$-orbital chain. 
In this function, there are two possible processes for the creation of a fermion in the $a$-orbital chain: (I) creation with an empty $b$ orbital [Fig.~\ref{fig4}(f)] and (II) creation with an occupied $b$ orbital [Fig.~\ref{fig4}(g)]. 
To account for these processes, we decompose the creation operator as 
\begin{align}
\hat{c}^{\dag}_{j,a} 
= \hat{c}^{\dag}_{j,a} \left( 1- \hat{n}_{j,b} \right) + \hat{c}^{\dag}_{j,a} \hat{n}_{j,b} 
= \hat{s}^{\dag}_{j,a} + \hat{d}^{\dag}_{j,a}, 
\end{align}
where $\hat{s}^{\dag}_{j,a}$ corresponds to the creation of a singly occupied site [case~(I), Fig.~\ref{fig4}(f)] while $\hat{d}^{\dag}_{j,a}$ represents the creation of a doubly occupied site [case~(II), Fig.~\ref{fig4}(g)]. 
Here, we demonstrate the propagation dynamics of an $a$-orbital fermion using the operators $\hat{s}^{\dag}_{j,a}$ and $\hat{d}^{\dag}_{j,a}$. 

Figures~\ref{fig4}(a)--\ref{fig4}(c) show the absolute values of the space-time correlation functions $\mathcal{G}_a(x,t) = -i\braket{\psi_0 | \hat{\mathcal{O}}_{j+x,a}(t) \hat{\mathcal{O}}^{\dag}_{j,a}(0) | \psi_0}$ for $\hat{\mathcal{O}}^{\dag}_{j,a} = \hat{c}^{\dag}_{j,a}$, $\hat{s}^{\dag}_{j,a}$, and $\hat{d}^{\dag}_{j,a}$. 
Here, $U=8t_{\rm h}$ is used to differentiate the energy scale of the doping-induced branch from others. 
In the large-$U$ regime, the correlators with different components, e.g., $\braket{\psi_0 | \hat{s}_{j+x,a}(t) \hat{d}^{\dag}_{j,a}(0) | \psi_0}$, give minor contributions. 
As seen in Fig.~\ref{fig4}(c), the broad propagation feature in Fig.~\ref{fig4}(a) is reproduced by $\hat{d}^{\dag}_{j,a}$. 
However, the slow propagation line, which is absent in Fig.~\ref{fig4}(c), arises from the $\hat{s}^{\dag}_{j,a}$ contribution as shown in Fig.~\ref{fig4}(b). 
In Figs.~\ref{fig4}(d) and \ref{fig4}(e), the single-particle spectrum $A_{a}(k,\omega)$ is compared with the spectral function obtained by $\mathcal{G}_a(x,t)$ associated with $\hat{s}^{\dag}_{j,a}$. 
As shown in Fig.~\ref{fig4}(e), the spectrum for $\hat{\mathcal{O}}^{\dag}_{j,a} = \hat{s}^{\dag}_{j,a}$ only constructs the low-energy branch near $E_{\rm F}$. 
Although it is a partial contribution of the spectral weights in $A_a(k,\omega)$, the essential spectral structure of the in-gap branch near $E_{\rm F}$ is mainly constructed by the contribution of $\hat{s}^{\dag}_{j,a}$. 
On the other hand, since $\hat{d}^{\dag}_{j,a}$ creates a fermion with the energy increase of $U$ [see Fig.~\ref{fig4}(g)], the upper band away from $E_{\rm F}$ in Fig.~\ref{fig4}(d) is mainly attributed to the dynamical properties described by $\hat{d}^{\dag}_{j,a}$. 
Therefore, the propagation of a fermion created by case (I) [Fig.~\ref{fig4}(f)] causes the emergence of the doping-induced branch near $E_{\rm F}$. 

\begin{figure}[t]
\begin{center}
\includegraphics[width=\columnwidth]{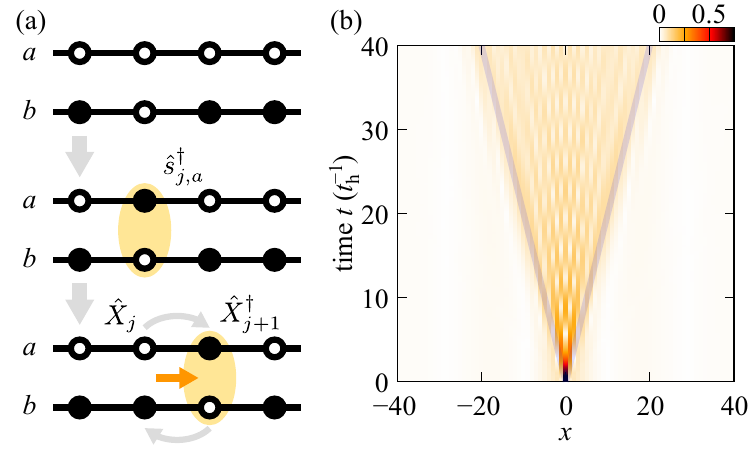} 
\caption{(a) Schematic figure of the single-particle creation and the propagation of the created particle along the $a$-orbital chain. 
(b) Absolute value of the correlation function $\mathcal{G}_X(x,t) = -i \braket{\psi_0 | \hat{X}_{j+x}(t) \hat{X}^{\dag}_{j}(0) | \psi_0}$ for $\hat{X}^{\dag}_j = \hat{c}^{\dag}_{j,a} \hat{c}_{j,b}$, where $U=8t_{\rm h}$, $D=0.92t_{\rm h}$, and $\delta=0.05$. 
The translucent blue lines in (b) represent $x=\pm J_{\rm ex}t$, where $J_{\rm ex}=4t_at_b/U$.} 
\label{fig5}
\end{center}
\end{figure}

Figure~\ref{fig5}(a) schematically shows the propagation dynamics of a fermion created by $\hat{s}^{\dag}_{j,a}$. 
Since the ground state is hole-doped, there are unoccupied sites, in which both $a$ and $b$ orbitals are empty [top panel in Fig.~\ref{fig5}(a)]. 
The operator $\hat{s}^{\dag}_{j,a} = \hat{c}^{\dag}_{j,a} (1-\hat{n}_{j,b})$ creates a fermion on an $a$ orbital when the other orbital is empty [middle panel in Fig.~\ref{fig5}(a)]. 
Then, the created fermion begins to propagate along the $a$-orbital chain. 
However, the motion of the created fermion is not simply allowed by the free hopping $t_a$ due to the Coulomb repulsion $U$. 
Particularly, in the large-$U$ case ($U \gg t_a, t_b$), the motion of a fermion along the $a$-orbital chain is strongly coupled to the motion of a hole along the $b$-orbital chain to avoid the creation of a doubly occupied site. 
In this case, the propagation of the created fermion is allowed by the particle exchange process shown in the bottom panel of Fig.~\ref{fig5}(a). 
This second-order process is characterized by the energy of $J_{\rm ex} = 4 t_a t_b /U$. 
When the creation operator of a local exciton is defined as $\hat{X}^{\dag}_j = \hat{c}^{\dag}_{j,a} \hat{c}_{j,b}$, the second-order particle exchange process is described by $J_{\rm ex} \hat{X}^{\dag}_{j+1} \hat{X}_j$. 
Hence, the particle propagation along the $a$-orbital chain can be characterized by the excitonic operator $\hat{X}_j$. 
To demonstrate its dynamics, we presents the space-time correlation function $\mathcal{G}_X(x,t) = -i\braket{\psi_0 | \hat{X}_{j+x}(t) \hat{X}^{\dag}_{j}(0) | \psi_0}$ in Fig.~\ref{fig5}(b). 
The time scale of the excitonic propagation in Fig.~\ref{fig5}(b) is consistent with that for the $a$-orbital fermion shown in Fig.~\ref{fig4}(b). 
The lines in Fig.~\ref{fig5}(b), which represent the slope of the propagation of $\mathcal{G}_X(x,t)$, are characterized by $J_{\rm ex}$. 
The slope occurring in $\mathcal{G}_X(x,t)$ [Fig.~\ref{fig5}(b)] is in good agreement with that for the fermion propagation in Fig.~\ref{fig4}(b). 
This correspondence indicates that the doping-induced branch, which is obtained by the Fourier transformation of $\mathcal{G}_a(x,t)$ in Fig.~\ref{fig4}(b), reflects the excitonic correlation.


\section{Summary and Outlook} \label{sec:summary}

We investigated the spectral properties of doped EIs. 
Employing the matrix-product-state-based methods, we computed the spectral functions for the correlated two-band model. 
We found the emergence of the doping-induced in-gap branch in the single-particle spectrum. 
This spectral property is distinct from that of conventional band insulators, in which carrier doping is achieved without deforming their band structures. 
In the optical conductivity, the spectral structure at $\omega>0$ of the nondoped state is maintained, while the Drude weight is generated with doping. 
We examined the origin of the doping-induced branch using the projected fermion operators. 
We showed that the doping-induced branch is strongly associated with the excitonic correlation. 
Our findings suggest that the doping-induced branch can serve as an indicator of the electron-hole correlation. 

In this paper, we used the hopping amplitude $t_{\rm h}$ as the unit of energy. 
If we set $t_{\rm h} \sim 0.3$~eV, the calculated $\sigma(\omega)$ for the nondoped state in Fig.~\ref{fig3} exhibits a similar spectrum to the optical conductivity of Ta$_2$NiSe$_5$~\cite{YLu2017,TLarkin2017}. 
A more quantitative discussion of Ta$_2$NiSe$_5$, which undergoes a structural transition, requires a spectral evaluation that incorporates electron-lattice coupling because our calculation only considered the electron-electron interaction. 
Electron-lattice coupling may also influence the occurrence of the doping-induced branch. 
Carrier doping in bulk samples has been realized by elemental substitution, such as (Ta$_{1-x}$Ti$_x$)$_2$NiSe$_5$~\cite{STsuchida2025}. 
We expect that the spectral properties of doped EI candidates will be uncovered in the future. 
Our model assumed that the Coulomb interaction is unaffected by doping, whereas changes in Coulomb interactions due to carrier doping potentially occur. 
It is also quantitatively important to consider screening effects on Coulomb interactions when discussing doping in real materials. 
Since our model [Eq.~\eqref{eq:Hamiltonian}] is spinless, the pseudospin representation for orbitals $a$ and $b$ maps our two-orbital model to the Hubbard model with spin-dependent hoppings under a Zeeman field. 
It is therefore natural that the spectral structure changes upon doping, as observed in doped Mott insulators. 
When spin degrees of freedom are active~\cite{JKunes2014_1,JNasu2016,DGeffroy2019}, richer branches incorporating spin dynamics can also emerge in doped EIs. 
These issues are interesting topics for future research.


\begin{acknowledgments}
We thank K.~Aido, M.~Akiyama, J.~Han, Y.~Inokuma, T.~Kondo, K.~Kuroki, R.~Mizuno, M.~Ochi, Y.~\={O}no, and S.~Tsuchida for fruitful discussions. 
This work was supported by Grants-in-Aid for Scientific Research from JSPS, KAKENHI Grant No.~JP24K06939, No.~JP24H00191, and No.~JP24K01333.
R.U. was supported by the Program for Leading Graduate Schools: ``Interactive Materials Science Cadet Program'' and JST SPRING, Grant No.~JPMJSP2138. 
This project was made possible by the DLR Quantum Computing Initiative and the Federal Ministry for Economic Affairs and Climate Action; qci.dlr.de/projects/ALQU. 
S.E. acknowledges the Q-Neko project, which has received funding from the European Union's Horizon Europe research and innovation programme under Grant Agreement No. 101241875. 
This work was also performed for Council for Science, Technology and Innovation (CSTI), Cross-ministerial Strategic Innovation Promotion Program (SIP), ``Promoting the application of advanced quantum technology platforms to social issues'' (Funding agency: QST). 
The DMRG and TEBD simulations were performed using the ITensor library~\cite{ITensor,ITensor2}. 
\end{acknowledgments}


\section*{Data Availability}
The data that support the findings of this article are openly available~\cite{data}.


\bibliography{reference}

\begin{thebibliography}{75}
\bibitem [{Emi()}]{Emin}%
  \BibitemOpen
  \href@noop {} {}\bibinfo {note} {$E_{\rm min}(N \pm 1) = \min\{E_0(N_a \pm
  1,N_b),E_0(N_a,N_b \pm 1)\}$, where $E_0(N_a,N_b)$ is the lowest energy in
  the $(N_a,N_b)$ sector.}\BibitemShut {Stop}%
\end{thebibliography}

\end{document}